# Perdeuteration of poly[2-methoxy-5-(2'-ethylhexyloxy)-1,4-phenylenevinylene] (d-MEHPPV): control of microscopic charge-carrier spin-spin coupling and of magnetic-field effects in optoelectronic devices†

Dani M. Stoltzfus,[a‡] Gajadhar Joshi,[b‡] Henna Popli,[b] Shirin Jamali,[b] Marzieh Kavand,[b] Sebastian Milster,[c] Tobias Grünbaum,[c] Sebastian Bange,[c] Adnan Nahlawi,[c] Mandefro Y. Teferi,[b] Sabastian I. Atwood,[b] Anna E. Leung,[d] Tamim A. Darwish,[d] Hans Malissa,*[b] Paul L. Burn,*[a] John M. Lupton*[bc] and Christoph Boehme*[b]

Control of the effective local hyperfine fields in a conjugated polymer, poly[2-methoxy-5-(2'-ethylhexyloxy)-1,4-phenylenevinylene] (MEHPPV), by isotopic engineering is reported. These fields, evident as a frequency-independent line broadening mechanism in electrically detected magnetic resonance spectroscopy (EDMR), originate from the unresolved hyperfine coupling between the electronic spin of charge carrier pairs and the nuclear spins of surrounding hydrogen isotopes. The room temperature study of effects caused by complete deuteration of this polymer through magnetoresistance, magnetoelectroluminescence, coherent pulsed and multi-frequency EDMR, as well as inverse spin-Hall effect measurements, confirm the weak hyperfine broadening of charge carrier magnetic resonance lines. As a consequence, we can resolve coherent charge-carrier spin-beating, allowing for direct measurements of the magnitude of electronic spin-spin interactions. In addition, the weak hyperfine coupling allows us to resolve substantial spin-orbit coupling effects in EDMR spectra, even at low magnetic field strengths. These results illustrate the dramatic influence of hyperfine fields on the spin physics of organic light-emitting diode (OLED) materials at room temperature, and point to routes to reaching exotic ultra-strong resonant-drive regimes needed for the study of light-matter interactions.

## Introduction

Spin-dependent recombination processes in organic semiconductors, such as in organic light-emitting diodes (OLEDs), are governed by the interplay between the weak but non-zero spin-orbit coupling (SOC), and the hyperfine coupling between charge carrier spins and surrounding nuclear spins. In OLEDs made of conjugated polymers the proportion of the structure of the materials made of protons is relatively high due to the need for solubilizing groups, and the unresolved hyperfine coupling between the electronic spin system and a large number of remote nuclear spins manifests itself as effective, randomly varying magnetic fields, i.e., the hyperfine fields.[1] Since the nuclear spin ensemble does not polarize thermally under the experimental conditions achievable, the magnitude of these fields is largely frequency-independent, and appears as an additional inhomogeneous line broadening mechanism in magnetic resonance spectroscopy, as most extensively studied recently using electrically detected magnetic resonance (EDMR) spectroscopy.[2-4] This broadening is observed in particular at low microwave (MW) excitation frequencies, where other, frequency-dependent line broadening mechanisms—such as field-dependent broadening due to shifts in the g-factor arising from SOC—become negligible.[2,3,5] The width of the local hyperfine field distribution throughout a macroscopic material limits the degree of coherent spin control by resonant MW excitation, and we find that materials with weak hyperfine fields not only exhibit narrow resonance lines but also more pronounced Rabi oscillations due to the longer $T_2^*$ dephasing times, which are a measure of the effective magnetic field inhomogeneity experienced by the precessing spins.[6-9]

Control over the magnitude of hyperfine fields can be achieved by isotopic engineering, e.g., replacement of protium with another deuterium. It was previously observed that the partial deuteration of a conjugated polymer, poly[2-methoxy-5-(2'-ethylhexyloxy)-1,4-phenylenevinylene] (MEHPPV), namely the 2-ethylhexyloxy side-chain, and related PPV derivatives, leads to significant changes in the magnetoresistance (i.e., the change of

[a.] Centre for Organic Photonics & Electronics, School of Chemistry and Molecular Biosciences, The University of Queensland, Brisbane, Queensland 4072, Australia.
[b.] Department of Physics and Astronomy, University of Utah, 115 South 1400 East, Salt Lake City, UT 84112, USA.
[c.] Institut für Experimentelle und Angewandte Physik, Universität Regensburg, Universitätsstraße 31, 93053 Regensburg, Germany.
[d.] National Deuteration Facility, Australian Nuclear Science and Technology Organization (ANSTO), Locked Bag 2001, Kirrawee DC, NSW 2232, Australia.
† Electronic Supplementary Information (ESI) available.
‡ These authors contributed equally to this work.

device conductivity as a function of magnetic field) as well as the EDMR characteristics,[1,10] which are both governed by SOC and the hyperfine fields.[11] In order to examine this effect more closely, we have synthesized perdeuterated MEHPPV (Sections S2,S3 ESI†), where all protium atoms (side-chains and polymer backbone) are substantially replaced by deuterium (d-MEHPPV). Such perdeuteration of a conjugated polymer constitutes a challenge because many of the starting materials are not available commercially in deuterated form and hence have to be prepared by deuteration of the protonated equivalents. We performed a detailed EDMR study of the new d-MEHPPV incorporated in OLED structures (Sections S4,S5, ESI†). We compare the results with commercial-grade protonated MEHPPV (h-MEHPPV) with a natural mixture of isotopes (i.e., mostly protium), which has been characterized in detail in previous studies.[1,3,5,6,10,12-14] Both materials are nominally structurally identical but differ in the hydrogen isotope composition and molecular-weight distribution, with the latter a result of the polymerization method which is difficult to control precisely.[15] We therefore probe directly the changes to spin-dependent device current that originate from the nuclear spin ensemble and its interaction with the electronic spin system.[16,17]

## Experimental results

The synthesis and characterization procedure for the preparation of the d-MEHPPV is described in the ESI†. Fig. 1 shows the magnetoresistance (red) and magneto-electroluminescence (magnetoEL, blue) of a d-MEHPPV OLED at room temperature as a function of magnetic field from -25 mT to +25 mT. The plotted values are normalized to the corresponding steady-state values at zero magnetic field. We observe a broad magnetic-field response that extends well beyond the measurement range and is described accurately by the phenomenological model (solid line) discussed in Joshi et al.[11] In addition, we observe an ultra-small magnetic field effect on the scale below 1 mT, which we do not discuss further here.

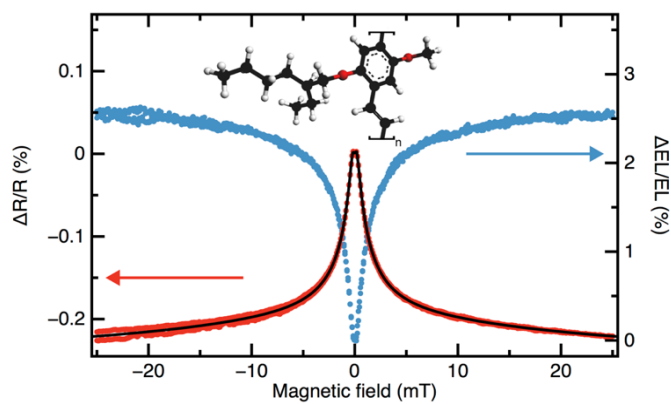

Figure 1 - Magnetoresistance (red) and magneto-electroluminescence (magnetoEL, blue) of OLEDs made of d-MEHPPV, along with a least-squares fit to the model described in Joshi et al.[11] (black line). Note that the measurements were performed under constant current to minimize the impact of conductivity changes on EL. The chemical structure of poly[2-(2-ethylhexyloxy-$d_{17}$)-5-methoxy-$d_3$-1,4-phenylenevinylene-$d_4$] (d-MEHPPV) is shown in the inset.

Note that the measurements were performed at constant current so that the magnetoEL is not directly controlled by the magnetoresistance. Next, we consider the effect of deuteration on dynamic spin-dependent recombination effects. Fig. 2 shows a multi-frequency EDMR analysis of d-MEHPPV OLED devices, i.e., continuous-wave EDMR spectra measured at several different excitation frequencies of up to 20 GHz.[3] The spectra measured are shown as individual lines in Fig. 2a. The spectra are described by a superposition of two Gaussian lines,[18] accounting for the inhomogeneous broadening of the resonances of the two pair partners (electron and hole), and are measured under conditions of magnetic field modulation, at a modulation frequency of 500 Hz and a modulation amplitude of approximately 0.15 mT. A pronounced broadening of the EDMR spectra is observed for frequencies above 100 MHz in Fig. 2b. Comparable data for h-MEHPPV have been previously published elsewhere and did not show frequency-dependent spectral broadening effects below 1 GHz.[3]

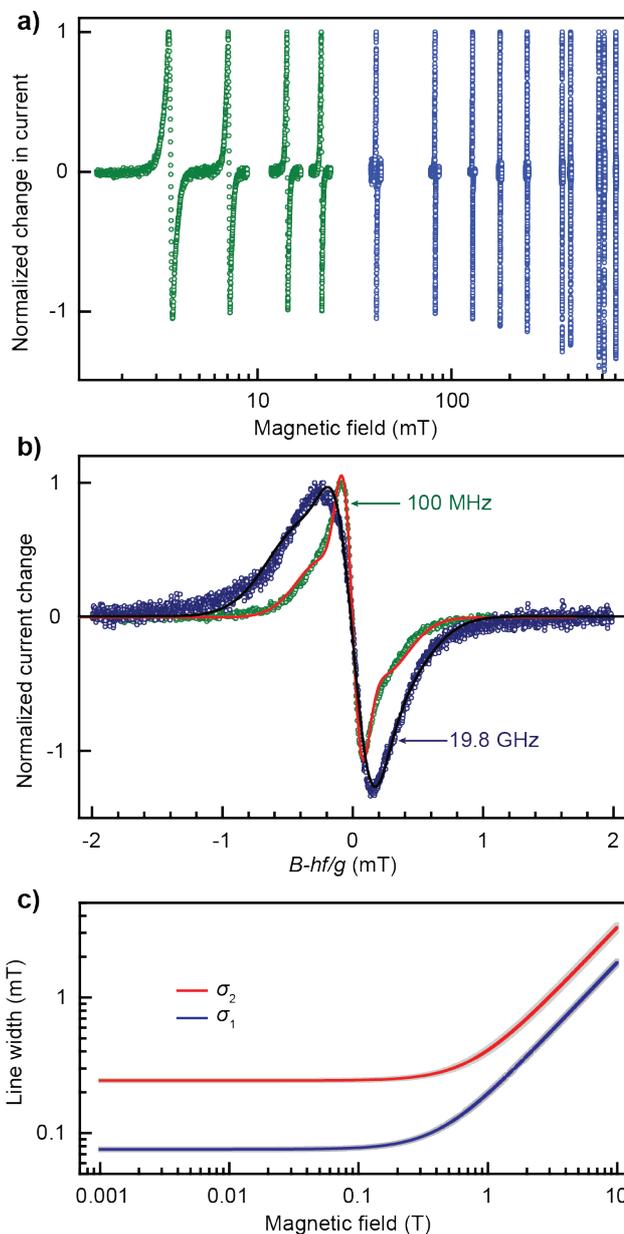

Figure 2 - Multi-frequency analysis of d-MEHPPV OLED devices at microwave frequencies below 20 GHz.[2,3,5,8,34] (a) Measured resonance lines normalized to the maximal current amplitude. (b) Comparison of the measured spectra at the lowest (100 MHz) and highest (19.8 GHz) frequencies, with the magnetic field axis normalized to the resonance field. The solid line represents the result of a global fit of all spectra obtained at all applied frequencies. (c) Variation of the root-mean-square resonance line widths of electron (red) and hole (blue) spins, as a function of the applied magnetic field (i.e., the excitation frequency). The displayed functions are the result of the numerical multi-frequency line-shape analysis of the data shown in (a) and (b), and the narrow grey shaded areas around the red and blue plots indicate the uncertainties in the line width obtained from a bootstrap analysis as described in Joshi et al.[3]

To assess the effect of deuteration on decoherence of the spin excitation, which is measured in the OLED device current controlled by spin-dependent electron-hole recombination, we recorded electrically detected spin-echoes using a Hahn echo sequence adapted for EDMR as explained in detail previously.[1,13,19-21] The π/2-τ-π-τ-π/2-echo pulse sequence used is illustrated in Fig. 3a: a resonant π/2-pulse (16 ns duration) is applied in order to rotate the spin packets from the thermal equilibrium orientation along the static magnetic field $B_0$ into a plane perpendicular to $B_0$. The spin packets dephase rapidly (with a time constant of $T_2^*$) during a waiting time τ. After the time τ, a 32 ns long π-pulse is applied to refocus the spin packets, which leads to the formation of the echo signal at time 2τ. In contrast to a conventional Hahn echo sequence, we applied another π/2-pulse following the two-pulse Hahn-echo state, resulting in electrically detectable spin-dependent recombination currents. The spin decoherence time, $T_2$, was determined by varying the delay time 2τ and assessing the exponential decay of the echo amplitude as a function of 2τ as shown in Fig. 3b.

Next, we examine the characteristics of coherent spin precession during pulsed EDMR experiments. Fig. 4 shows measurements of Rabi spin-beat oscillations as a function of external magnetic field, plotted as a change in integrated device current as a function of MW pulse length. During the application of the MW pulse, the spins of the pair precess, giving rise to a sequence in order to project the spin packet onto the direction of $B_0$ during the echo and, thus, onto a permutation symmetry sinusoidal oscillation in spin permutation symmetry. Since there are two spins present, either one of the two spins of the pair—electron or hole—may precess with a Rabi frequency of $\Omega_R$, yet for weak magnetic resonant MW driving fields smaller than the hyperfine fields, $B_1 < \Delta B_{\text{hyp}}$, it is unlikely that both pair partners will be on magnetic resonance at the same time. For sufficiently strong drive conditions, however, both of the spins may precess at once, leading to a doubling in the Rabi frequency. In this situation, spin beating, which is also referred to as spin locking, occurs. The coherent oscillations, either fundamental or harmonic, between eigenstates of the pair with increasing pulse duration give rise to sinusoidal oscillations in the integrated device current signal. By gradually detuning the MW pulse from resonance, which is achieved by sweeping the magnetic field across resonance, the excitation spectrum of the pulse is set to overlap with either a single pair partner or with both pair partners, leading to oscillations at the fundamental frequency and at the second harmonic. A detailed explanation of the measurement procedure, including a theoretical treatment of the phenomenon, is provided in van Schooten et al.[9] Fig. 4a,b show examples of the integrated current transients as a function of pulse duration for the case of off-resonance (blue) and on-resonance (red) excitation of OLEDs comprising d-MEHPPV. The magnetic field values for which Fig. 4a,b were measured are indicated by blue and red arrows in the two-dimensional color plot in Fig. 4c. As the magnetic field is swept, following Rabi's frequency formula, the precession frequency of both the fundamental and the second-harmonic oscillation changes.

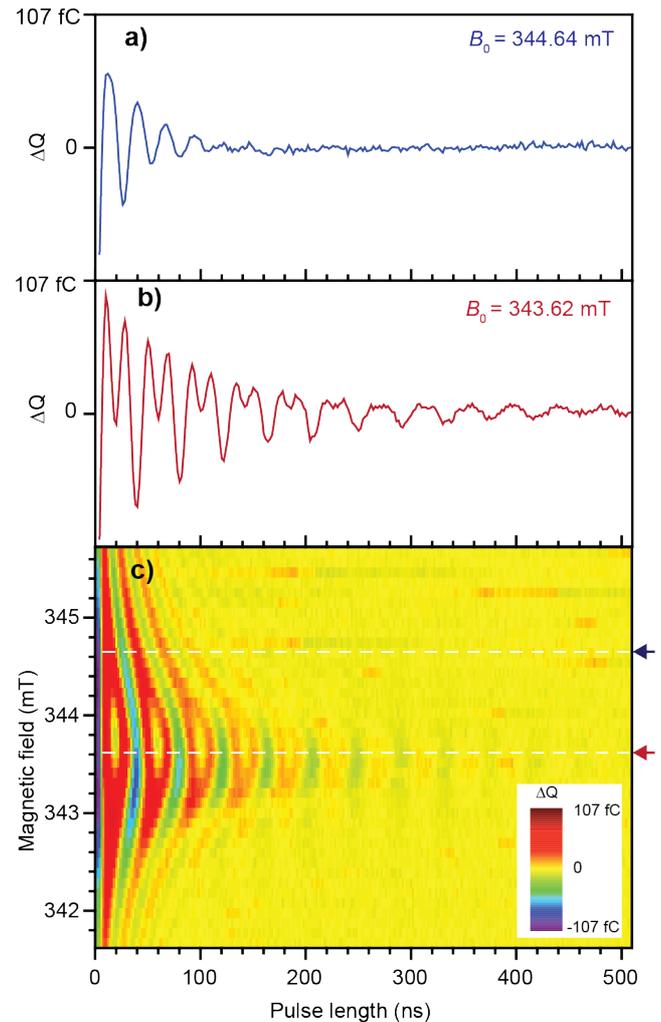

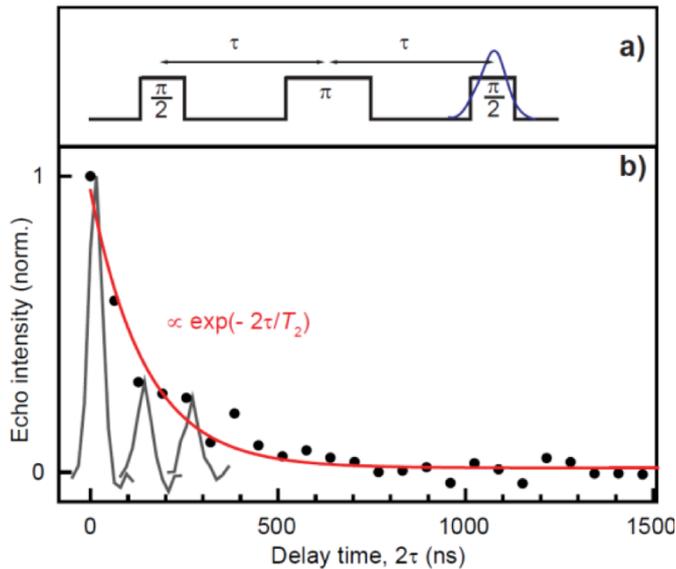

Figure 3 - Determination of the spin dephasing time $T_2$ in d-MEHPPV by an electrically detected Hahn echo measurement.[1,13,19-21] (a) Illustration of the modified Hahn echo pulse sequence for electrical detection: π/2-τ-π-τ-π/2. The final π/2-pulse is not part of the conventional Hahn echo pulse sequence and serves to project the rephased spin echo onto a field axis parallel to the external field to enable electrical detection through a current measurement. The length of a π-pulse in these experiments was 32 ns. (b) Plots of the current-detected echo amplitude determined from individual measurements as a function of the delay time 2τ. The red solid line represents a least-squares fit to an exponential decay time $T_2$. Three examples of measured current echoes are superimposed such that their echo maxima overlap with the data point of the corresponding values of 2τ (grey curves).

Figure 4 - Electrically detected charge-carrier spin-Rabi oscillations and evidence for spin-beating in d-MEHPPV OLEDs in a static magnetic field range close to the electron paramagnetic resonance condition under application of a ~9.7 GHz radiation pulse (i.e., at X-band). (a) Integration of the device current change from its steady-state value (i.e., the charge ΔQ) as a function of the length of the applied MW pulse for an arbitrary, slightly off-magnetic resonance magnetic field ($B_0$ = 344.64 mT). The oscillation of the charge signal with pulse duration is due to a coherently driven Rabi nutation between the eigenstates of the spin pairs. (b) Data recorded from an experiment similar to (a) with identical experimental conditions, except for the static magnetic field, which is set to the resonance maximum ($B_0$ = 343.62 mT). Here, both charge carrier spins within the charge carrier pair nutate with time, so that a quantum beat (spin-locking) signal at the second harmonic frequency of the fundamental oscillation period emerges. (c) Contour plot of the Rabi oscillations shown in (a) and (b) as a function of the external magnetic field, showing the detuning from magnetic resonance. The blue and red arrows indicate the magnetic fields that correspond to the conditions of the data in panels (a) and (b).

Finally, Fig. 5 shows measurements of the inverse spin-Hall effect (ISHE) current at Cu contacts on a d-MEHPPV film adjacent to a ferromagnetic (FM) NiFe layer. During FM resonance (FMR), spin pumping is achieved by the magnetization dynamics of the NiFe film that generates a pure spin current in the d-MEHPPV layer.[22-24] The ISHE leads to the conversion of the spin current into an electromotive force due to the SOC in d-MEHPPV, which is detected as a charge current between the two lateral Cu contacts of the device. The inset of Fig. 5 shows the device structure, which is described in detail in Sun et al.[22] Crucially, the experiments are performed under pulsed MW irradiation so as to avoid heating artifacts due to the ferromagnetic resonance. Fig. 5 shows the ISHE current, measured as a function of the applied static magnetic field for two opposing in-plane magnetic field orientations [0° (solid circles) and 180° (open circles)] with the proximal NiFe thin film driven in ferromagnetic resonance. In both cases, the current reaches an extremum of approximately 140 nA at a magnetic field value that corresponds to the resonance condition of the NiFe film for that particular orientation. These orientations correspond to the condition for which spin-pumping from the ferromagnet into the organic semiconductor occurs.[22] The strength of the driving field, $B_1$, was measured independently as 0.615 mT at an orientation of 0° and 0.595 mT at 180° for a MW pulse power of 1000 W,[25] which is the maximum nominal output power of the travelling-wave tube amplifier used for these experiments.

## Discussion

The magnetoresistance response for d-MEHPPV is narrower compared to that of h-MEHPPV. Fig. 1 shows a least-squares fit of an empirical function to the measured data set to describe the magnetoresistance response (solid black line).[11] The line shape is dominated by an expression of the form $f(B) = $ const. $+ [B/(|B| + b)]^2$,[11,26-28] with the fitting parameter $b$ = 0.433 mT, implying that the magnetoresistance response of d-MEHPPV is approximately eight times narrower than in regular protonated h-MEHPPV and still significantly narrower than in the previously studied partially deuterated MEHPPV in which the 2-ethylhexyloxy side-chain was deuterated.[§,10,27,29-33] For the multi-frequency analysis of d-MEHPPV in Fig. 2, we performed a global nonlinear least-squares fit of the data.[2,3,5,34] The fitting procedure considers a line-shape model as a function of both the static magnetic field $B_0$ and the excitation frequency. The line shape—a superposition of two Gaussian lines with different g-factors (i.e., resonance positions) and line widths assigned to the two charge-carrier spins of electron and hole—takes frequency-dependent (e.g., SOC-related) and frequency-independent contributions (i.e., due to disorder in the local magnetic-field strength arising from inhomogeneity in the external field as well as the hyperfine-field distributions) into account.[18] The solid lines in Fig. 2b and the frequency dependence of the resulting overall line widths of the two constituent peaks of the resonance as shown in Fig. 2c are obtained directly from this global least-squares fit. The fitted line widths exhibit a plateau in the low-field limit that corresponds to the hyperfine field strength experienced by the two charge carriers of the pair. The widths increase linearly with magnetic field strength in the limit of higher fields, where SOC effects dominate—these are the distribution in g-factors, i.e. the Δg-effect, g-strain broadening, and the influence of anisotropic g-tensors. The gray shaded areas in Fig. 2c represent the 95% confidence intervals of the extracted line width values, obtained by the statistical bootstrap analysis, which is a numerical procedure to establish fit parameter confidence intervals directly from the noise in the experimental datasets.[5,35] These values allow an assessment of the relative contributions of hyperfine coupling and SOC to line broadening. The high reliability of this fit is demonstrated by the extraordinarily narrow range of the fit results: the uncertainty ranges marked in grey are barely discernible in Fig. 2c. This high degree of reliability arises due to the simultaneous (global) fit of all measured spectra.

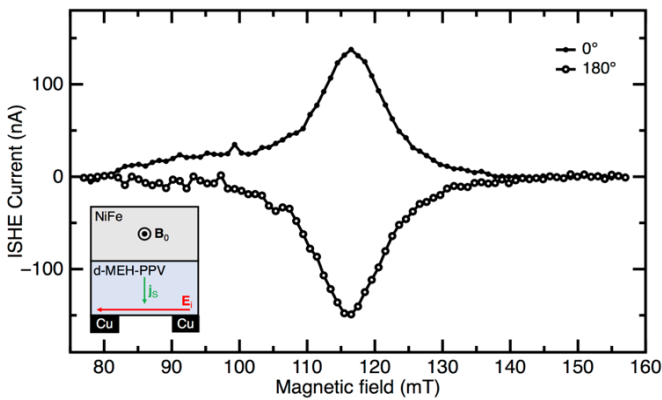

Figure 5 - Pulsed inverse spin-Hall (ISHE) response as a function of magnetic field, measured for opposing in-plane magnetic field orientations (0° and 180°). In brief, a spin current is generated in the organic semiconductor by the ferromagnetic resonance driven in the NiFe top contact. The inset shows a sketch of the device, consisting of a layer of d-MEHPPV on top of Cu electrodes, and covered with a NiFe film. An electric field $E_i$ is generated between the two Cu electrodes and the orthogonal current $j_s$ due to the ISHE is measured. The relative orientations of the static magnetic field, the injected spin current, and the resulting electromotive force on the charges are described in Sun et al.[22] The ISHE provides a metric for the strength of spin-orbit coupling in the semiconductor material.

From the global non-linear least-squares fit of the multi-frequency datasets, we obtain line-shape parameters for the narrow line (labelled as line 1, which is associated with the hole spin) and the broad line (line 2, associated with the electron spin).[5] Each line is represented by (i) an isotropic g-factor; (ii) a frequency-independent line width $\Delta B_{1/2}$; and (iii) a g-strain parameter $\Delta g$ that describes the frequency-dependent line broadening due to SOC.[§§] The line shape parameters obtained are $g_1 = 2.003318 \pm 0.000006$, $\Delta B_{1/2}^{(1)} = (0.178 \pm 0.001)$ mT, and $\Delta g_1 = (0.854 \pm 0.014) \times 10^{-3}$, for resonance line 1, and $g_2 = 2.003542 \pm 0.000009$, $\Delta B_{1/2}^{(2)} = (0.574 \pm 0.004)$ mT, and $\Delta g_2 = (1.549 \pm 0.046) \times 10^{-3}$ for line 2. The uncertainties that originate from the bootstrap analysis are (cf. Fig. 2c) small compared to the analogous values found in h-MEHPPV.[3] The saturation FWHM line widths at low fields, i.e., the static inhomogeneous broadening effects due to the hyperfine fields, are substantially smaller than the corresponding values found for h-MEHPPV of $(0.5595 \pm 0.0007)$ mT and $(1.7018 \pm 0.0061)$ mT.[3] The ratios between these values (2.51 for line 1 and 2.96 for line 2) reflect the much lower hyperfine fields present in the d-MEHPPV compared to commercial h-MEHPPV. The reduced hyperfine coupling arises due to the lower nuclear magnetic moment of $^2$H compared to $^1$H, which has an abundance of 99.98% in the natural isotope composition in conventional h-MEHPPV.[6,7]

Fig. 2c shows the variation of the root-mean-square (RMS) line widths $\sigma_{1,2}$ with magnetic field. Phenomenologically, the functional dependence of $\sigma_{1,2}$ is given as $\sigma^2 = (\Delta B_{\text{hyp}})^2 + \alpha^2 B_0^2$, where $\alpha$ is the frequency-dependent line broadening parameter.[3] Both parameters can be expressed as a FWHM line width $\Delta B_{1/2} = \sqrt{2\ln 2}\, \Delta B_{\text{hyp}}$ and the g-strain parameter $\Delta g$, which is related to $\alpha$ by $\Delta g = \sqrt{2\ln 2}\, \alpha g$. Both nomenclatures are completely equivalent.[5] Here, the parameters $\alpha$ are determined as $\alpha_1 = 1.81 \times 10^{-4}$ and $\alpha_2 = 3.28 \times 10^{-4}$, which are similar to the values reported previously for commercial h-MEHPPV, $\alpha_1 = 1.78 \times 10^{-4}$ and $\alpha_2 = 4.82 \times 10^{-4}$. Note that these parameters depend on the strength of SOC in the material and are not directly influenced by the nuclear isotope species. The slight differences in the apparent influence of SOC between the two materials, which implies a slightly larger SOC in the d-MEHPPV, could arise from differences in molecular weight and dispersity of the two polymers (h-MEHPPV: $M_w = 3.8 \times 10^5$, Đ = 4.7; d-MEHPPV: $M_w = 4.2 \times 10^5$, Đ = 3.3) leading to subtle differences in local chain conformation and hence film morphology and/or density.

The red solid line in Fig. 3b shows the result of a least-squares fit of an exponential decay to the measured echo amplitudes as a function of pulse delay time $\tau$. From the time constant of this exponential decay, we can directly evaluate the spin-dephasing time $T_2$. From the data in Fig. 3, we obtain a value of $T_2 = (146 \pm 11)$ ns, which, is considerably shorter than the spin dephasing time of 348 ns that has been reported for conventional h-MEHPPV at room temperature.[13] This difference is not straightforward to interpret because the phase-memory time measured here is influenced not only by spin-spin relaxation processes, but also by spectral diffusion, spin diffusion, and instantaneous diffusion that originates from random spin flips of other non-resonant dipole-coupled charge-carrier spins.[36] The details of each process may differ slightly in both materials, most likely due to small local differences in the polymer conformation and hence film morphology and/or density. It is also conceivable that the acceleration in spin dephasing relates to the slight increase in the effect of SOC observed in the frequency-dependent spectra in Fig. 2 that can be corroborated through the ISHE measurements discussed below.

Having established the intrinsic decoherence time of the spins, we now turn to the dephasing phenomena determined by the time $T_2^*$. Fig. 6 shows a Fourier analysis of the Rabi spin-beat oscillations in the device current plotted in Fig. 4. Depending on the value of the applied static magnetic field $B_0$, i.e., the degree of detuning from magnetic resonance, and thus the spectral overlap of the excitation spectrum of the pulse with either one or both charge-carrier spins, different frequency components are exhibited by the Rabi oscillations detected. These frequency values follow Rabi's frequency formula as a function of detuning, and fundamental frequency components as well as the beat components—which arise due to simultaneous precession of electron and hole spins—are observed.[9] Fig. 6a,b show the frequency spectrum for the two examples of off-resonance and on-resonance excitation from Fig. 4, and Fig. 6c shows a colour plot of the variation of these frequency spectra with magnetic field. Due to the weaker hyperfine field distributions in d-MEHPPV, and in contrast to the case of conventional h-MEHPPV,[9] these frequency components are pronounced and discernible even for larger detuning of the magnetic field.

The observation of spin-beating in the Rabi flopping through the device current allows for estimates of the zero-field charge-carrier spin coupling strengths within charge-carrier pairs. These pairs—intermediate precursor pairs that eventually transition to strongly bound excitons—are Coulombically bound and experience a coupling interaction determined by dipolar and exchange mechanisms.[9,37-39] Fig. 7a shows the precession frequency maxima, extracted from the Fourier transform in Fig. 6 as a function of $B_0$ for both the fundamental frequency and the second-harmonic precession component. The solid line shows the analytical solution for the $B_0$-dependence of the second-harmonic precession frequency in the limit of weak spin-spin coupling.[9] A small but significant deviation between the measured and calculated values is visible from plotting the difference between these two values in Fig. 7b. Evidently, the second-harmonic precession frequency is not precisely at twice the fundamental frequency. This difference depends on the exchange coupling strengths $J$ and the dipolar coupling strength $D$ between the two charge-carrier spins that form an intermediate pair. At the resonance center, i.e., at zero detuning, this difference is 656±297 kHz for d-MEHPPV. Note that in h-MEHPPV, in contrast, the resonance spectra are too broad to resolve the effect of detuning off-resonance on the beat oscillation.[9]

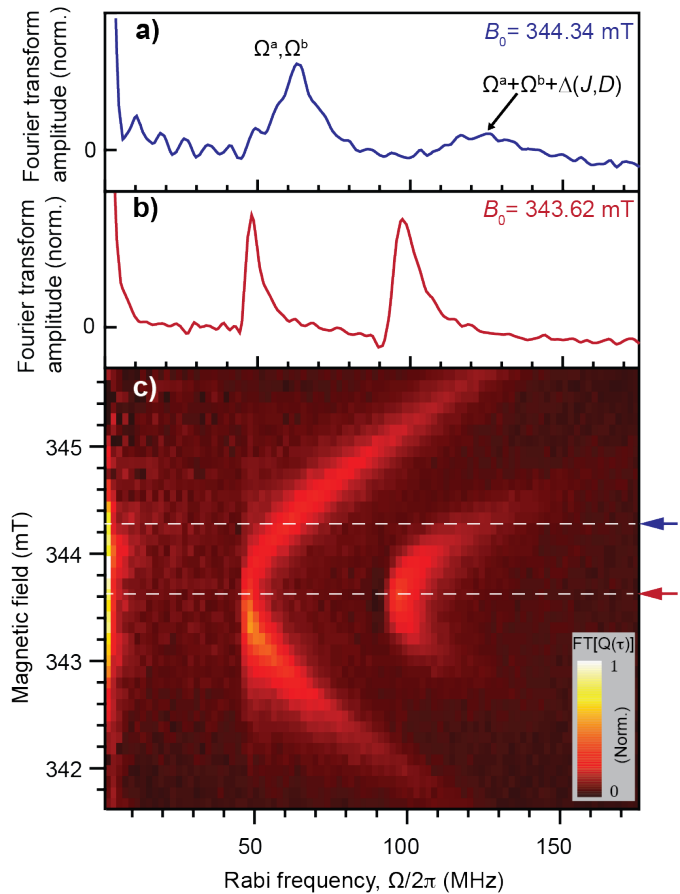
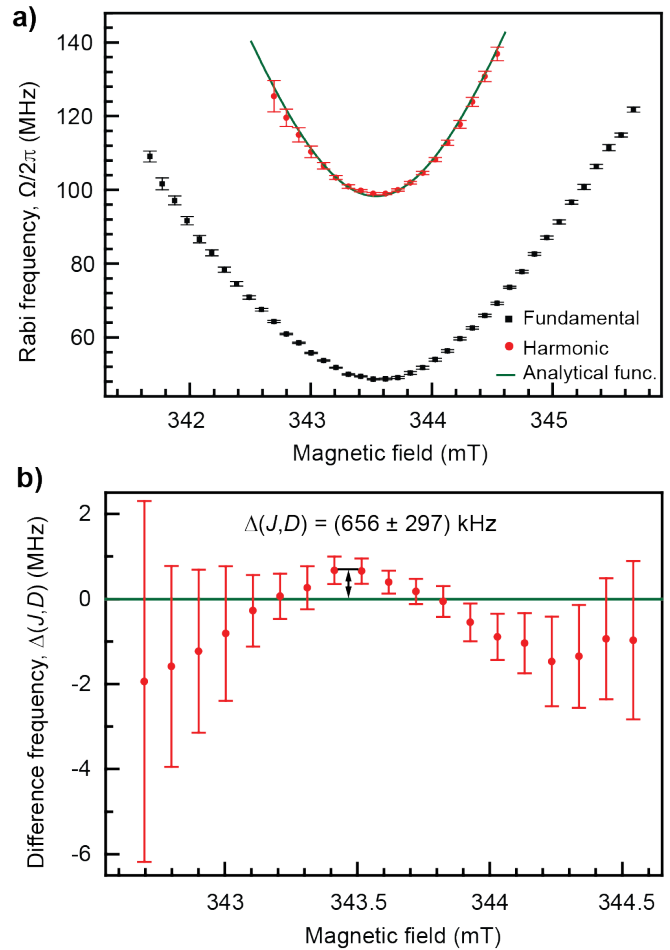

Figure 6 - Fourier analysis of the electrically detected spin-Rabi beating in d-MEHPPV as a function of static magnetic field strength shown if Fig. 4, as performed with the procedure described in van Schooten et al.[9] (a) and (b) Plots of the frequency spectra of the electrically detected spin-Rabi oscillations off- and on-resonance corresponding to the data sets displayed in panels (a) and (b) of Fig. 4. These panels reveal that spin-beating, indicated by the peak at a frequency $\Omega_a + \Omega_b$, is strong for on-resonance magnetic fields and significantly weaker when off-resonance. The beat frequency depends on the zero-field splitting $\Delta$ of the pair, which arises from exchange $J$ and dipolar $D$ spin-spin interactions. (c) Two-dimensional plot of the frequency spectrum over the entire detuning range of magnetic field strength over which the measurements were made. In contrast to the case of conventional h-MEHPPV,[9] the individual frequency components are well separated in d-MEHPPV.

Figure 7 - Estimation of zero-field splitting parameters of intermediate carrier paris in d-MEHPPV. (a) Plots of the fundamental and second-harmonic spin-beat frequencies extracted from the maxima of the frequency components shown in Fig. 6c, as functions of detuning off-resonance, i.e., in dependence of the applied static magnetic field strength. The solid line shows the analytical function describing the second-harmonic nutation, computed from the fundamental nutation frequency as described by Rabi's formula.[9,37-39] (b) Plot of the difference between the measured frequency of the spin-beat component in the Rabi flopping shown in (a) and the frequency computed for spin pairs with negligible zero-field splitting. The difference provides an estimate of the zero-field splitting frequency $\Delta$ of the spin pair. Since $\Delta$ arises from exchange coupling $J$ as well as dipole-dipole coupling $D$ within the intermediate paris, this value allows bounds to be set for both values of $J$ and $D$. $\Delta$ therefore provides insight into the geometric size of the electron-hole pairs probed in EDMR.

The ISHE response from d-MEHPPV shown in Fig. 5 is approximately 2.5 times stronger than the ISHE current measured in an analogous device in which a similar 1,4-phenylenevinylene-derived polymer, poly(di-n-octyloxy-1,4-phenylenevinylene) (DOO-PPV) was used.[22] It is also approximately twice as strong compared to the ISHE signal in devices based on the polymer blend material PEDOT:PSS and is comparable to the signal strength found in Pt-Q, a polymer that consists of intrachain Pt atoms separated by units of 5,8-diethynyl-2,3-diphenylquinoxaline. This latter material is known to possess charge-carrier states with enhanced SOC strengths due to the presence of Pt atoms with high atomic order number. However, the ISHE signal strength observed with d-MEHPPV is still almost five times weaker compared to that of $C_{60}$-based devices. These latter structures consist exclusively of carbon with low atomic order number but have electronic states with enhanced SOC due to their mixing with non-bonding orbitals arising from the curvature of the molecules.[22] The observed increase of the ISHE signal of d-MEHPPV and, thus, the inferred increase in SOC compared to h-MEHPPV and other π-conjugated polymers, appears to be in line with the observed increase of the frequency-dependent spectral broadening of Fig. 3, which is also attributed to SOC. We conclude that d-MEHPPV appears to show somewhat stronger SOC than the conventional h-MEHPPV. Since there is no apparent reason why the degree of SOC in d-MEHPPV should depend directly on the hyperfine interaction between nuclear and charge-carrier spins, we attribute this observed increase in SOC in d-MEHPPV to the subtle differences in the film morphology and/or density, which could arise from small local conformational differences along the polymer backbones.

## Conclusions

The study of spin-dependent electronic transitions and spin transport in a novel conjugated-polymer material, perdeuterated MEHPPV, reveals that the absence of proton nuclear spins that couple to the charge-carrier spins leads to a significant reduction in the random unresolved local hyperfine fields, causing a strong narrowing of charge-carrier spin resonance lines, in particular in the limit of low static magnetic fields. From the comparison of the underlying spin spectroscopy of d-MEHPPV with conventional protonated h-MEHPPV we conclude that randomly varying, slowly fluctuating hyperfine fields give rise to frequency-independent inhomogeneous line broadening which can obscure the macroscopic effects of spin coherence, such as Rabi spin-beating oscillations under coherent time-dependent microwave drive. Interestingly, even though such coherent oscillations are only resolved in d-MEHPPV and not in h-MEHPPV, the actual spin coherence times $T_2$ are not found to increase for the deuterated material—quite the opposite, they even appear to be slightly shorter. Nevertheless, the data presented here shows that materials with weak intrinsic hyperfine fields represent much better candidates to examine coherent spin-motion effects on the magneto-electronic behaviour of materials, and because of this, they open up new perspectives for high-sensitivity magnetometry.[4,8,9] We tentatively attribute the small increase of SOC-induced materials effects in d-MEHPPV compared to h-MEHPPV—the increased line broadening at high resonance frequencies and the stronger ISHE signal—to small differences in film morphology, which could potentially arise from subtle differences in local chain conformations. Such an increased SOC could also be responsible for the shortened coherence times in d-MEHPPV. This noted, however, there may be a more fundamental relation between reduced hyperfine coupling and enhanced SOC which we cannot pinpoint at present.

Finally, we note that, besides applications in magnetometry,[4] deuteration of organic semiconductor materials is particularly interesting for reaching unconventional magnetic-resonance drive conditions, where the Rabi frequency becomes comparable—or even exceeds—the carrier frequency in the ultra-strong drive regime.[10,40-42] The latter is determined by the Zeeman splitting of the spin levels, but is limited by the magnetic disorder of the system: the lower the overall hyperfine field strength, the lower the RF power necessary to transition the system into the ultra-strong drive regime. Therefore, we anticipate that with the availability of d-MEHPPV, the physics of new magnetic resonant drive regimes will be become accessible, in particular with regards to collective spin phenomena such as spin-Dicke states that emerge under ultra-strong drive.[10]

## Conflicts of interest

There are no conflicts to declare.

## Acknowledgements

This work was supported by the US Department of Energy, Office of Basic Energy Sciences, Division of Materials Sciences and Engineering under Award #DE-SC0000909. Magnetoresistance and magneto-electroluminescence measurements were conducted with support of the Sonderforschungsbereich 1277, Project B03, of the German Science Foundation.
The National Deuteration Facility is partly funded by NCRIS, an Australian Government initiative. PLB is an Australian Research Council Laureate Fellow and the work was supported in part by this Fellowship (FL160100067). The ISHE experiments were supported by NSF-DMR Award #1701427.

## Notes and references

Supporting Information

**Perdeuteration of poly[2-methoxy-5-(2'-ethylhexyloxy)-1,4-phenylenevinylene] (d-MEHPPV): control of microscopic charge-carrier spin-spin coupling and of magnetic-field effects in optoelectronic devices**

*Dani M. Stoltzfus, Gajadhar Joshi, Henna Popli, Shirin Jamali, Marzieh Kavand, Sebastian Milster, Tobias Grünbaum, Sebastian Bange, Adnan Nahlawi, Mandefro Y. Teferi, Sabastian I. Atwood, Anna E. Leung, Tamim A. Darwish, Hans Malissa\*, Paul L. Burn\*, John M. Lupton\*, and Christoph Boehme\**

**S1. Supporting figures**

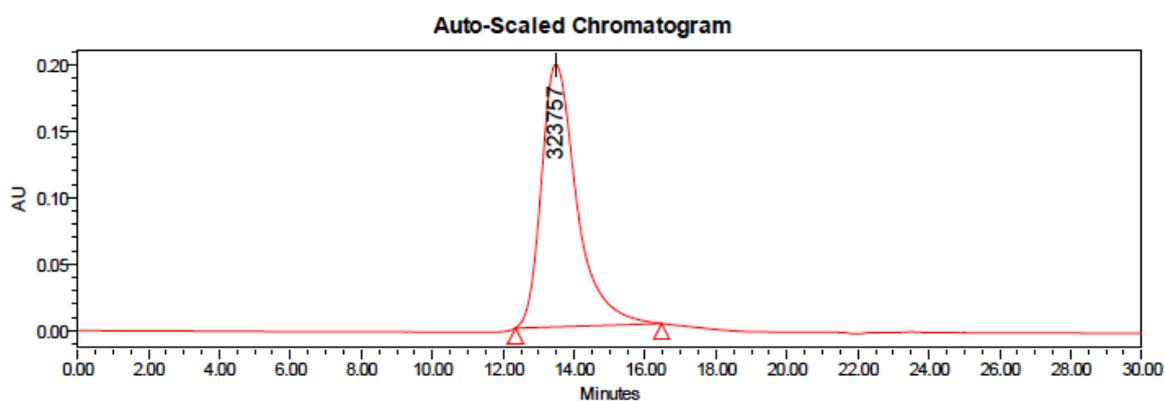

**Fig. S1** GPC trace of d-MEHPPV.

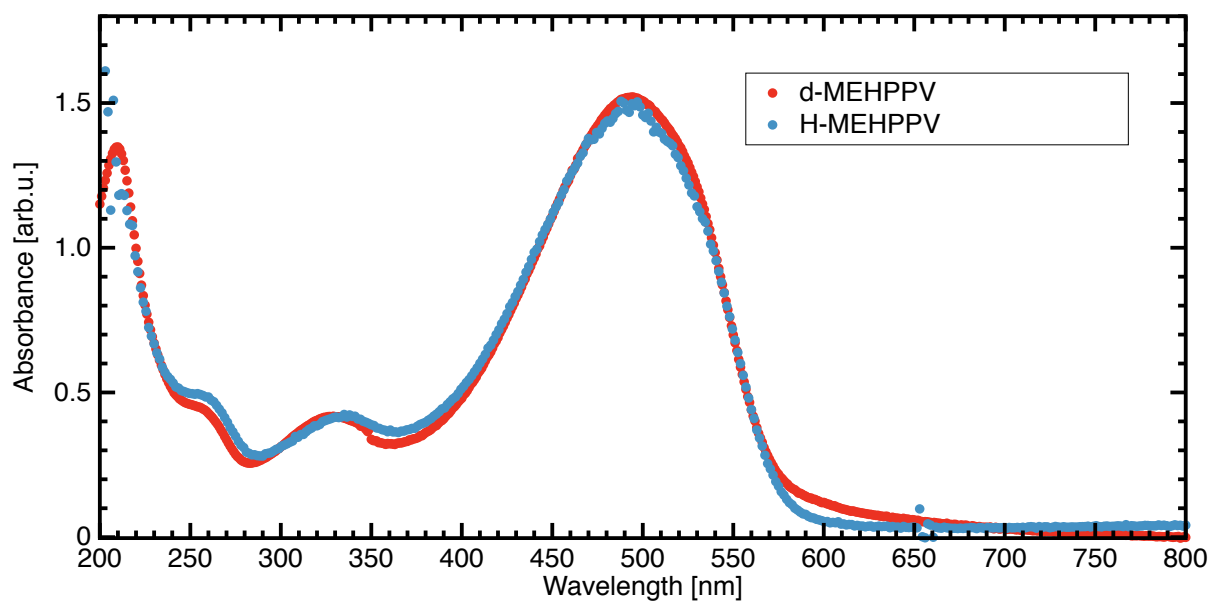

**Fig. S2** Film UV-visible spectrum of d-MEHPPV and h-MEHPPV. The localized (at ≈210 nm) to the delocalized (at ≈500 nm) π-π* transitions for both materials are essentially the same indicating that they have similarly delocalized chromophores.

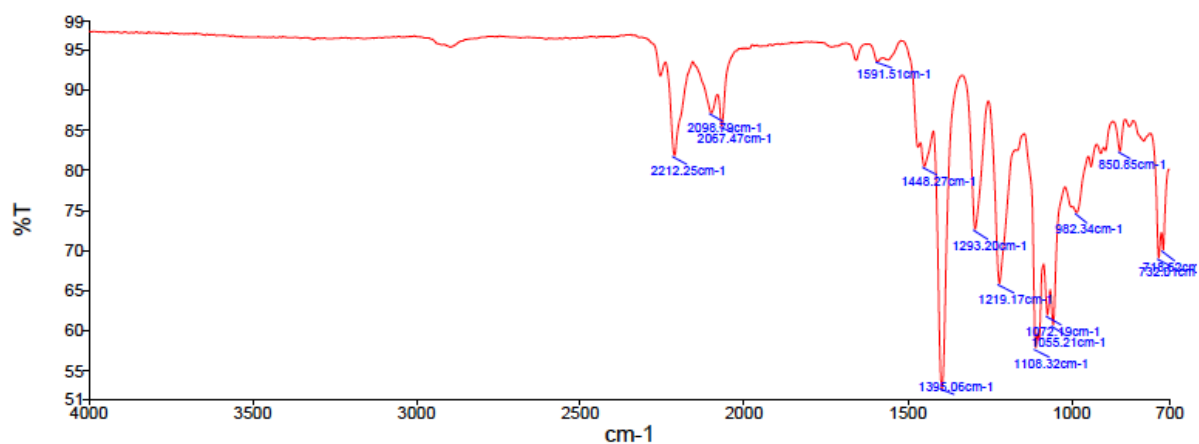

**Fig. S3** Infrared spectrum of d-MEHPPV.

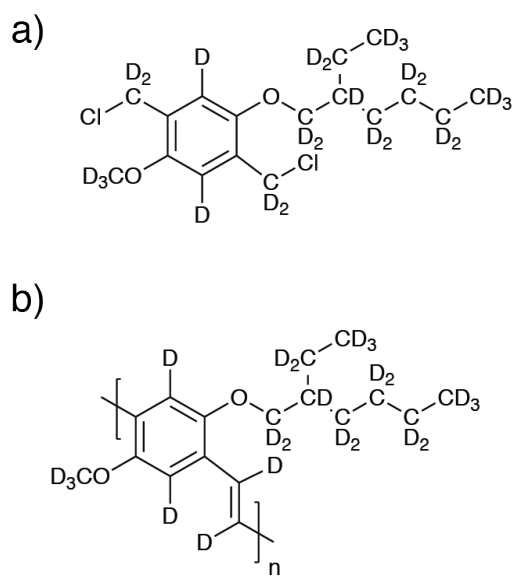

**Fig. S4** a) Monomer: 1,4-bis[chloromethyl]-2-[(2-ethylhexyl)oxy]-5-methoxybenzene-d$_{26}$ b) Polymer: poly[2-(2-ethylhexyloxy-d$_{17}$)-5-methoxy-d$_3$-1,4-phenylenevinylene-d$_4$] (d-MEHPPV).

## S2. Synthesis of d-MEHPPV

Monomer: Chemicals, including paraformaldehyde-$d_2$ (98% isotopic purity), iodomethane-$d_3$ (99.5% isotopic purity) and LiAlD$_4$ (98% isotopic purity) were used as received from Sigma-Aldrich. Solvents were used as received from Sigma Aldrich or were dried following literature methods. NMR spectroscopy solvents were used as received from Cambridge Isotope Laboratories Inc. D$_2$O (99.8%) was purchased from AECL, Canada. $^1$H NMR (400 MHz), $^2$H NMR (61.4 MHz) and $^{13}$C NMR (100 MHz) spectra were recorded on a Bruker 400 MHz spectrometer at 298 K. Chemical shifts were referenced to the residual signal of the solvent. $^2$H NMR spectroscopy was performed using the lock channel of the probe for direct observation. Electrospray ionization mass spectra were recorded on a 4000 QTrap AB SCIEX mass spectrometer. The overall deuteration of 2-ethylhexanoic acid was calculated by ER MS (enhanced resolution-MS) using the isotope distribution of the different isotopologues by analyzing the area under each MS peak, which corresponds to a defined number of deuterium atoms. The contribution of the carbon 13 (natural abundance) to the value of the area under each [X+1] MS signal was subtracted based on the relative amount found in the unlabelled version. The deuteration of hydroquinone was calculated using $^{13}$C NMR spectroscopy according to the method reported by Darwish et al.[S1]

Polymer: All reagents were used as received unless otherwise stated. Tetrahydrofuran was dried on an LC systems solvent purification system prior to use. UV-visible spectrophotometry was performed using a Cary 5000 UV-VIS spectrophotometer. The thin film of the polymer was spin-coated onto a fused silica substrate from chlorobenzene. FT-IR spectroscopy was performed on a solid sample using a Perkin-Elmer Spectrum 100 FT-IR spectrometer with an ATR attachment. Gel Permeation Chromatography (GPC) of the polymer was carried out on a Waters GPC 1515 system equipped with Empower software. The instrument was connected to a refractive index (RI) and an UV-vis detector, and the two columns [Styragel HT-3 and Styragel HT-6E (300 mm + 300 mm lengths, 7.8 mm diameter)] were kept at 40°C with a flow rate of 1 mL/min. No flow marker was used during the analysis. Narrow polystyrene standards in the $M_w$ range of 1350 Da to 1300000 Da were used to create a calibration curve. The sample was prepared in tetrahydrofuran at a concentration of 1 mg/mL and filtered through a 0.45 micron PTFE filter before injection.

*Synthesis of monomer 1,4-bis[chloromethyl]-2-[(2-ethylhexyl)oxy]-5-methoxybenzene-d$_{26}$*: the structure is shown in Fig. S4a.

*Hydroquinone-d4*: Hydroquinone (10.19 g, 92.54 mmol), NaOD (40% wt. in D$_2$O, 5.65 mL, 80.5 mmol), D$_2$O (120 mL), Pt/C (10% wt., 200 mg) and Pd/C (10% wt., 200 mg) were placed into a 600 mL Parr reactor, which was purged with nitrogen whilst stirring, then hydrogen whilst stirring, and then sealed. The vessel was heated to 150°C (maximum pressure observed: 4 bar) for 24 h, then the vessel was cooled and the contents removed. The reaction mixture was filtered through Celite to remove the catalysts and the filtrate was acidified to pH 2 with aqueous sulfuric acid (1 M). The mixture was extracted into diethyl ether (3×150 ml) and the combined organic extracts were filtered through a short silica plug using ether as eluent. The filtrate was collected and the solvent removed to provide an orange solid, which was suspended in cold ethylacetate:petroleum ether (3:7) and then filtered through a Buchner funnel to afford an off-white solid. The filtrate was collected and the solvent removed with the solid re-suspended in cold ethylacetate:petroleum ether (3:7). The mixture was filtered through a Buchner funnel to provide a second crop of an off-white solid, which was combined with the first to give hydroquinone-d$_4$ (6.01 g, 53%). $^1$H NMR (DMSO-d$_6$, 400 MHz) δ 6.55 (s, residual), 8.60 (s, OH). $^2$H NMR (DMSO-d$_6$, 61.4 MHz) δ 6.58 (s, 4 D). $^{13}$C NMR {$^1$H-decoupled} (DMSO-d$_6$, 100 MHz) δ 115.3 (m), 149.6 (s). $^{13}$C NMR {$^1$H and $^2$H-decoupled} (DMSO-d$_6$, 100 MHz) δ 115.3 (s), 149.6 (s).

*4-Methoxyphenol-d7*: A mixture of potassium carbonate (2.67 g, 19.3 mmol) and hydroquinone-d$_4$ (2.00 g, 17.5 mmol) in anhydrous acetone (50 mL) was stirred at room temperature for 1.5 h before methyliodide-d$_3$ (1.38 mL, 22.2 mmol) was added dropwise. The mixture was stirred for 2 h at room temperature and then potassium carbonate (702 mg, 5.08 mmol) and methyliodide-d$_3$ (970 μL, 16.0 mmol) (dropwise) were added sequentially. The mixture was allowed to stir at room temperature overnight and then filtered. The filtrate was collected and the solvent removed before the residue was dissolved in water (20 ml). The solution was acidified to pH 1 with hydrochloric acid (1 M) before being extracted with ethylacetate (3×100 mL). The combined organic extracts were filtered through a short silica plug using ethylacetate as eluent. The filtrate was collected and the solvent removed. The residue was purified by flash column chromatography over silica using ethylacetate:petroleum ether mixtures (1:4 to 3:7) as eluent, with the fractions containing the product visualized with iodine to afford 4-methoxyphenol-d$_7$ as a yellow crystalline solid (725 mg, 32%). $^1$H NMR (CDCl$_3$, 400 MHz) δ 3.73 (s, residual), 4.59 (br s, OH), 6.77 (s, residual), 6.79 (s, residual).

$^2$H NMR (CDCl$_3$, 61.4 MHz) δ 3.74 (s, 3 D), 6.83 (s, 4 D). $^{13}$C NMR {$^1$H-decoupled} (CDCl$_3$, 100 MHz) δ 55.1 (m), 114.6 (m), 115.8 (m), 149.5 (s), 153.8 (s). $^{13}$C NMR {$^1$H and $^2$H-decoupled} (CDCl$_3$, 100 MHz) δ 55.1, 114.6, 115.8, 149.5, 153.8.

*2-Ethylhexanoic acid-d$_{15}$*: to each of two 600 mL Parr reactors was added 2-ethylhexanoic acid (13.0 g, 90.1 mmol), D$_2$O (120 mL), NaOD (40% wt. in D$_2$O, 7.00 mL, 99.7 mmol) and Pt/C (10% wt., 400 mg). The vessels were purged with nitrogen whilst stirring, then sealed and heated to 220°C for 3 d (maximum pressure observed: 22 bar). The vessels were cooled and the contents were filtered through Celite to remove the catalyst. The filtrate was collected and acidified to pH 2 with hydrochloric acid (1 M). The aqueous mixture was then extracted with ethylacetate (3×200 mL) and the combined organic extracts were dried over anhydrous sodium sulfate, filtered, and the solvent removed to afford 2-ethylhexanoic acid-d$_{15}$ (27.04 g, 90.5% D by MS, 95%). The material was subjected to a second deuteration cycle as follows: to each of two 600 mL Parr reactors were added 2-ethylhexanoic acid-d$_{15}$ (13.50 g, 90.5% D by MS, 85.3 mmol), D$_2$O (120 mL), NaOD 40% wt. in D$_2$O, 6.60 ml, 94.0 mmol) and Pt/C (10% wt., 400 mg). The vessels were purged with nitrogen whilst stirring, then sealed and heated to 220°C for 3 d (maximum pressure observed: 22 bar). The vessels were cooled and the contents were filtered through Celite to remove the catalyst. The filtrate was acidified to pH 2 with hydrochloric acid (1 M) then extracted with ethylacetate (3×200 mL). The combined organic extracts were dried over anhydrous sodium sulfate, filtered and the solvent removed to afford 2-ethylhexanoic acid-d$_{15}$ (25.8 g, 96.1% D by MS, 96%). $^1$H NMR (CDCl$_3$, 400 MHz) δ 0.83 (s, residual), 0.88 (s, residual), 1.26 (s, residual), 1.43 (s, residual), 1.50 (s, residual), 1.58-1.59 (complex, residual), 2.25 (s, residual). $^2$H NMR (CDCl$_3$, 61.4 MHz) δ 0.84-0.90 (complex, 6 D), 1.25 (s, 4 D), 1.45-1.59 (complex, 4 D), 2.25 (s, 1 D). $^{13}$C NMR {$^1$H-decoupled} (CDCl$_3$, 100 MHz) δ 10.8 (m), 12.9 (m), 21.5 (m), 24.4 (m), 28.3 (m), 30.7 (m), 46.4 (m), 183.1 (s). $^{13}$C NMR {$^1$H and $^2$H-decoupled} (CDCl$_3$, 100 MHz) δ 10.9, 13.0, 21.6, 24.3, 28.4, 30.6, 46.5, 183.2. MS (ESI−) *m/z* calculated for C$_8$D$_{15}$O$_2$ [M−H]$^-$ as 158.2; found: 158.2. Deuteration: 96.1% by MS: isotope distribution: *d*$_{12}$ 2.0%, *d*$_{13}$ 8.6%, *d*$_{14}$ 34.0%, *d*$_{15}$ 55.4%.

*2-Ethylhexan-1-ol-d$_{17}$*: A solution of 2-ethylhexanoic acid-d$_{15}$ (13.00 g, 96.1% D by MS, 81.6 mmol) in dry tetrahydrofuran (50 mL) was added dropwise over 2 h to an ice-cold suspension of LiAlD$_4$ (4.70 g, 112 mmol) in dry tetrahydrofuran (150 mL) that had been placed under a nitrogen flow. When the addition was complete, the mixture was allowed to warm to room temperature before being heated at reflux overnight. The mixture was cooled in an ice bath and water was added slowly and cautiously to quench the remaining LiAlD$_4$. Aqueous sulfuric acid (1 M, 100 mL) was added and the mixture was extracted with diethyl ether (3×200 mL). The combined organic extracts were washed with saturated aqueous sodium hydrogen carbonate (150 mL), dried over anhydrous sodium sulfate, filtered, and the solvent removed to provide 2-ethylhexan-1-ol-d$_{17}$ as a clear oil (11.50 g, 96%), which required no further purification. $^1$H NMR (CDCl$_3$, 400 MHz) δ 0.83 (s, residual), 1.21-1.35 (complex, residual), 3.50 (s, residual). $^2$H NMR (CDCl$_3$, 61.4 MHz) δ 0.85 (s, 6 D), 1.22-1.53 (complex, 9 D), 3.52 (s, 2 D). $^{13}$C NMR {$^1$H-decoupled} (CDCl$_3$, 100 MHz) δ 10.0 (m), 13.1 (m), 21.9 (m), 22.2 (m), 27.9 (m), 29.0 (m), 40.9 (m), 64.6 (m). $^{13}$C NMR {$^1$H and $^2$H-decoupled} (CDCl$_3$, 100 MHz) δ 10.1, 13.0, 21.9, 22.2, 27.9, 29.0, 40.9, 64.6.

*3-(Bromomethyl)heptane-d$_{17}$ (2-ethylhexyl bromide-d$_{17}$)*: Triphenylphosphine (31.2 g, 119 mmol) was added to a solution of 2-ethylhexan-1-ol-d$_{17}$ (11.5 g, 78.1 mmol) in dry dichloromethane (180 mL). The mixture was stirred until the triphenylphosphine dissolved, and was then cooled in an ice bath. *N*-Bromosuccinimide (21.20 g, 118 mmol) was added portion-wise, allowing each portion to dissolve before the next was added. At the end of the addition, the bright yellow solution was stirred with ice bath cooling for 30 min, then allowed to warm to room temperature and stirred for an additional 4.5 h. The mixture was quenched with the addition of saturated aqueous sodium thiosulfate (100 mL). The layers were separated and the aqueous portion was extracted with dichloromethane (2×150 mL). The combined organic extracts were washed with brine (100 mL), then water (100 mL) before being filtered through a short silica plug using dichloromethane as eluent. The filtrate was collected and the solvent removed to leave a pink oil containing a suspension of a white solid. Petroleum ether was added to the mixture and the suspension was filtered through a second short silica plug using petroleum ether as eluent. The filtrate was collected and the solvent removed to afford 3-(bromomethyl)heptane-d$_{17}$ (13.88 g, 85%) as a clear oil. $^1$H NMR (CDCl$_3$, 400 MHz) δ 0.83-0.90 (complex, residual), 1.14-1.36 (complex, residual), 1.50 (s, residual), 3.41-3.42 (complex, residual). $^2$H NMR (CDCl$_3$, 61.4 MHz) δ 0.86 (s, 6 D), 1.19-1.37 (complex, 8 D), 1.50 (s, 1 D), 3.43 (s, 2 D). $^{13}$C NMR {$^1$H-decoupled} (CDCl$_3$, 100 MHz) δ 9.9 (m), 13.0 (m), 21.7 (m), 24.3 (m), 27.6 (m), 30.8 (m), 38.7 (m), 40.1 (m). $^{13}$C NMR {$^1$H and $^2$H-decoupled} (CDCl$_3$, 100 MHz) δ 9.9, 13.0, 21.7, 24.1, 27.6, 30.8, 38.7, 40.1.

*1-[(2-Ethylhexyl)oxy]-4-methoxybenzene-d$_{24}$*: 3-(Bromomethyl)heptane-d$_{17}$ (2.31 g, 11.0 mmol) was added to a solution of 4-methoxyphenol-d$_7$ (1.20 g, 9.15 mmol) in dry *N,N*-dimethylformamide (35 mL). Sodium tert-butoxide (1.77 g, 18.4 mmol) was added and the mixture was heated at 110°C overnight, then cooled. Saturated aqueous ammonium chloride (50 mL) was added. The mixture was extracted with dichloromethane (3×50 mL) and the combined organic extracts were washed with water (6×100 mL), dried over anhydrous sodium sulfate, filtered, and the solvent removed. The residue was purified by flash column chromatography over silica using a dichloromethane:petroleum ether mixture (1:4) as eluent (visualized with UV light) to provide 1-[(2-ethylhexyl)oxy]-4-methoxybenzene-d$_{24}$ as a clear oil (2.06 g, 86%). $^1$H NMR (CDCl$_3$, 400 MHz) δ 0.846-0.864 (complex, residual), 1.26-1.37 (complex, residual), 1.66 (s, residual), 3.73-3.76 (complex, residual), 6.82-6.84 (complex, residual). $^2$H NMR (CDCl$_3$, 61.4 MHz) δ 0.86-0.88 (complex, 6 D), 1.26-1.45 (complex, 8 D), 1.66 (s, 1 D), 3.75 (complex, 5 D), 6.88 (s, 4 D). $^{13}$C NMR {$^1$H-decoupled} (CDCl$_3$, 100 MHz) δ 10.1 (m), 13.1 (m), 21.9 (m), 22.7 (m), 27.9 (m), 29.4 (m), 38.4 (m), 55.1 (m), 70.5 (m), 114.3 (m), 115.2 (m), 153.6 (s), 153.7 (s). $^{13}$C NMR {$^1$H and $^2$H-decoupled} (CDCl$_3$, 100 MHz) δ 10.1, 13.1, 21.9, 22.7, 27.9, 29.4, 38.4, 55.1, 70.5, 114.3, 115.2, 153.59, 153.64.

*1,4-Bis[chloromethyl]-2-[(2-ethylhexyl)oxy]-5-methoxybenzene-d$_{26}$*: Concentrated hydrochloric acid (37%, 12.6 mL, 153 mmol) and then acetic anhydride (20.3 mL, 215 mmol) (dropwise as exothermic) were added to a mixture of 1-[(2-ethylhexyl)oxy]-4-methoxybenzene-d$_{24}$ (2.02 g, 7.75 mmol) and paraformaldehyde-d$_2$ (650 mg, 20.3 mmol). When the addition was complete, the mixture was heated to 80°C overnight, before being allowed to cool to room temperature and diluted with water (50 mL). The mixture was extracted with ethylacetate (3×50 mL) and the combined organic extracts were washed with brine (100 mL), then saturated aqueous sodium hydrogen carbonate (100 mL), dried over anhydrous sodium sulfate, filtered, and then the solvent was removed to afford an off-white solid. The residue was purified by flash column chromatography over silica using dichloromethane:petroleum ether mixtures (0:1 to 1:9) as eluent to provide 1,4-bis[chloromethyl]-2-[(2-ethylhexyl)oxy]-5-methoxybenzene-d$_{26}$ as a white solid (2.06 g, 74%). $^1$H NMR (CDCl$_3$, 400 MHz) δ 0.86-0.89 (complex, residual), 1.26-1.47 (complex, residual), 1.70 (s, residual), 3.82-3.85 (complex, residual), 4.63 (d, *J* = 1.00 Hz, residual), 6.85 (s, residual), 6.91-6.92 (complex, residual). $^2$H NMR (CDCl$_3$, 61.4 MHz) δ 0.86-0.90 (complex, 6 D), 1.28-1.70 (complex, 9 D), 3.84 (complex, 5 D), 4.63 (s, 4 D). $^{13}$C NMR {$^1$H-decoupled} (CDCl$_3$, 100 MHz) δ 10.4 (m), 13.0 (m), 21.9 (m), 22.9 (m), 27.9 (m), 29.5 (m),

38.5 (m), 41.0 (m), 55.6 (m), 70.6 (m), 113.4 (m), 114.1 (m), 126.7 (s), 126.9 (s), 150.97 (s), 151.02 (s). $^{13}$C NMR {$^{1}$H and $^{2}$H-decoupled} (CDCl$_3$, 100 MHz) δ 10.2, 13.0, 21.9, 22.9, 27.9, 29.5, 38.5, 41.0, 55.6, 70.4, 113.1, 113.8, 126.7, 126.9, 150.97, 151.01. The integration of the $^{2}$H NMR spectrum of the final compound, 1,4-bis[chloromethyl]-2-[(2-ethylhexyl)oxy]-5-methoxybenzene-d$_{26}$, demonstrates that the deuteration is consistent across the molecule, indicating that no D/H exchange occurred during any of the synthetic steps.

*Synthesis of polymer, poly[2-(2-ethylhexyloxy-d$_{17}$)-5-methoxy-d$_3$-1,4-phenylenevinylene-d$_4$] (d-MEHPPV)*: the structure is shown in Fig. S4b

*Poly[2-(2-ethylhexyloxy-d$_{17}$)-5-methoxy-d$_3$-1,4-phenylenevinylene-d$_4$] (d-MEHPPV)*: A solution of 1,4-bis[chloromethyl]-2-[(2-ethylhexyl)oxy]-5-methoxybenzene-d$_{26}$ (0.500 g, 1.40 mmol) in anhydrous tetrahydrofuran (40 mL) was stirred at room temperature under argon. Freshly sublimed potassium tert-butoxide (0.98 g, 8.79 mmol) in anhydrous tetrahydrofuran (10 mL) was added in one portion, and the solution was stirred for 4 h at room temperature. The resulting gel was diluted with tetrahydrofuran (~500 mL) and chloroform (~500 mL) until the material was fully dissolved, and then methanol (~250 mL) was added. The precipitate was collected via vacuum filtration and was dissolved in tetrahydrofuran (400 mL) before methanol (~600 mL) was added. The mixture was centrifuged (10 min, 2000 rpm) and the majority of the supernatant removed before the precipitate was collected via vacuum filtration. The precipitate was then dissolved in tetrahydrofuran (400 mL) and methanol (600 mL) was added. The mixture was centrifuged (10 min, 2000 rpm) and the majority of the supernatant was removed before the precipitate was collected via vacuum filtration and was dried *in vacuo* to afford poly[2-(2-ethylhexyloxy-d$_{17}$)-2-methoxy-d$_3$-1,4-phenylenevinylene-d$_4$] as a red powder (0.150 g, 36%); IR (solid) λ/cm$^{-1}$: 2212, 2099, 2067, 1448, 1395, 1293, 1219, 1108, 1072, 1055, 982, 851, 732, 719; λ$_{max}$(film)/nm 208, 256sh, 326, 491; GPC (THF, 40°C), $M_w$ = 4.2x10$^5$, $M_n$ = 1.3x10$^5$, Đ = 3.3. A GPC trace, UV-vis spectrum and an IR spectrum of d-MEHPPV are shown in Figs. S1-3.

The molecular weight and polydispersity of the conventional h-MEHPPV are $M_w$ = 3.8x10$^5$ and Đ = 4.7 as stated by the manufacturer.

**S3. Analysis of the degree of deuteration d-MEHPPV**

Based on the isotopic purities of LiAlD$_4$, methyliodide-d$_3$ and paraformaldehyde-d$_2$, and the percentage deuteration of hydroquinone-d$_4$ and 2-ethylhexanoic acid-d$_{15}$ it was calculated that the isotopic purity of the synthesized MEHPPV-d$_{24}$ would be 97±2%. The analysis was determined from the following data and assumptions:

1. Paraformaldehyde-d$_2$ (98% isotopic purity) was purchased from Sigma-Aldrich; the percentage deuteration of the vinylene moieties is thus 98%.

2. Iodomethane-d$_3$ (99.5% isotopic purity) was purchased from Sigma-Aldrich; the percentage deuteration at the methoxy methyl group is thus 99.5%.

3. The percentage deuteration at the aromatic positions was calculated having determined the deuteration of hydroquinone-d$_4$ via $^1$H and $^{13}$C NMR spectroscopy. The $^{13}$C {$^1$H, $^2$H-decoupled} NMR spectrum of hydroquinone-d$_4$ was used to determine the percentage deuteration to be 95.4±0.5%, using a comparison of the integration of the analogous carbon sites at both the quaternary and tertiary positions, according to Darwish *et al.*[S1] The quaternary sites are represented by two resonances at 149.733 and 149.684 ppm. The resonance at 149.684 ppm (arbitrary integration of 1) is assigned as the quaternary carbon flanked by a deuterated tertiary carbon on either side. The resonance at 149.733 ppm (integrating for 0.114) is assigned as the quaternary carbon flanked by one deuterated tertiary carbon and one protonated tertiary carbon. The percentage deuteration is thus calculated by:

$$(1 + 0.114 / 2) / (1 + 0.114) = 94.9\%$$

4. The tertiary sites are represented by three resonances at 115.617, 115.451 and 115.359 ppm. The resonance at 115.359 ppm is assigned at the deuterated tertiary carbon adjacent to another deuterated tertiary carbon, while the resonance at 115.451 ppm is assigned as the deuterated tertiary carbon adjacent to a protonated tertiary carbon (combined integration of 2.144). The resonance at 115.617 is assigned as the corresponding protonated tertiary carbon adjacent to a deuterated tertiary carbon (integration of 0.094). The percentage deuteration is thus calculated by:

$$2.144 / (2.144 + 0.094) = 95.8\%$$

5. The $^1$H NMR spectrum of hydroquinone-d$_4$ was used to determine that the minimum percentage deuteration is 94.2%, using a comparison of the integration of the four residual aromatic protons with the hydroxyl protons.

6. The percentage deuteration of the branched alkyl chain (except for the methylene group adjacent to the ether) was calculated having determined the deuteration of the starting acid, 2-ethylhexanoic acid-d$_{15}$, to be 96.1±2% via mass spectrometry.

7. LiAlD$_4$ (98% isotopic purity) was purchased from Sigma-Aldrich; the percentage deuteration at the methylene adjacent to the ether is thus 98%.

8. The overall deuteration was determined using steps 1-5 as follows:

$$(0.98 \times 2) + (0.995 \times 3) + (0.954 \times 2) + (0.961 \times 15) + (0.98 \times 2) = x \times 24$$
$$x = (1.96 + 2.985 + 1.908 + 14.415 + 1.96)/24$$
$$x = (23.228/24)$$
$$x = 96.8\%$$

## S4. Fabrication of OLEDs

d-MEHPPV solutions were prepared inside a glove box with a N$_2$ atmosphere in order to avoid contamination with O$_2$ and H$_2$O. The d-MEHPPV was dissolved in toluene at a concentration of 4.5 g/L. As the d-MEHPPV did not dissolve easily at room temperature, the solution was heated to 50-70°C on a hot plate while stirring until it had dissolved, after several days. The OLED devices were prepared by spin-coating of the d-MEHPPV solution on a previously deposited PEDOT:PSS/indium-tin-oxide stack on a glass substrate, as described previously.[S2,S3] The spin-coater was operated at 550 rpm, and a time delay of 55 s was introduced between the application of the d-MEHPPV onto the substrate and the spin-coating procedure to ensure wetting. Thermally deposited Al/Ca electrodes were used to ensure bipolar charge-carrier injection.

## S5. Experimental measurement procedures

Magnetoresistance and magneto-electroluminescence (magnetoEL) measurements were performed in a custom-made uncooled electromagnet powered by a CAEN ELS easy driver 5020 bipolar power supply. The samples were operated under a constant current of 100 μA using a Keithley 238 high-current source measure unit, and the change in device voltage was recorded as a function of the magnetic field and digitized with appropriate acquisition software. Note that since the devices are operated in constant current mode to be able to record the magneto-electroluminescence, the overall magnitude of magnetoresistance is smaller than that usually reported for conditions of constant voltage. The EL was collected using an optical fibre and directed onto a Femto OE-200 low-noise silicon photoreceiver, which was read out by a Keysight 34461A multimeter. We performed EDMR (using both continuous-wave excitation with magnetic field modulation and with ns-range pulsed excitation) on the same devices used for the magnetoresistance measurements in a commercial spectrometer (Bruker ElexSys 580). For X-band (~9.7 GHz) MW frequencies, we used a cylindrical dielectric resonator (Bruker FlexLine ER4118X-MD5) for magnetic resonant excitation. For other MW frequencies (between 100 MHz and 20 GHz) we used an Agilent EXG N5173B frequency generator with custom-designed EDMR probe-heads with coplanar waveguide resonators for frequencies between 1 GHz and 20 GHz, and NMR-style radiofrequency coils for frequencies between 100 MHz and 1 GHz.[S4,S5] In all cases, EDMR was recorded by applying a constant voltage with a battery source (Stanford Research Systems SIM928), and detecting the resulting current changes at magnetic resonance with a transimpedance amplifier (SRS 570) with adjustable frequency filters, a bandpass filter for the range 100 Hz to 3 kHz for continuous wave and 100 Hz to 100 kHz for pulsed experiments. For continuous-wave measurements, the output of the current amplifier was connected to the built-in lock-in amplifier of the Bruker Elexsys E580 facility, while for pulsed measurements, it was connected to the built-in digitizer. Electrical detection of the spin echo was made by integration of the transient spin-dependent device current recorded following the echo pulse sequence, over an interval of 15 μs. This integration was achieved using a boxcar integrator (SRS 250).[S6]

We also performed pulsed ISHE spectroscopy by measuring the electric current in the d-MEHPPV films under ferromagnetic resonant (FMR) excitation of an adjacent ferromagnetic

layer.[S7–S9] For these measurements, dedicated devices were prepared, consisting of a 240 nm thick d-MEHPPV layer that was located on top of two Cu electrodes and was covered with a NiFe film. FMR in the NiFe was excited by 2 µs long MW pulses at a power of 1000 W, which caused injection of a pure spin current into the d-MEHPPV layer for the duration of the pulse. In the polymer film, the spin current was then converted into an electromotive force through the ISHE, which in turn was detectable through a current measurement at the Cu electrodes. ISHE spectroscopy as well as details about the device structure are described in Sun *et al.*,[S7] and a sketch of the device architecture is shown in the inset of Figure 7. The exact MW power at the position of the sample was established independently from inductively detected Rabi oscillations on a separate spin standard (a 1:1 complex of α,γ-bisdiphenylene-β-phenylallyl and benzene, BDPA, a free radical).[S10] All of the above-described experiments were performed at room temperature.